\documentstyle[aps,12pt]{revtex}
\begin{document}
\draft
\title{Disordered Flat Phase in a Solid on Solid Model of Fcc(110) Surfaces and
Dimer States in Quantum Spin-1/2 Chains}
\vspace{10mm}
\author{Giuseppe Santoro}
\address{International School for Advanced Studies,
Via Beirut 4, 34014 Trieste, Italy}
\author{Michele Fabrizio}
\address{Institut Laue-Langevin,
Avenue des Martyres, BP166X, Grenoble, France}
\maketitle
\vspace{15mm}
\begin{abstract}
We present a restricted solid on solid hamiltonian for fcc (110) surfaces.
It is the simplest generalization of the exactly solvable BCSOS model
which is able to describe a $(2\times 1)$ missing-row reconstructed surface.
We study this model by mapping it onto a quantum spin-1/2 chain of the
Heisenberg type, with second and third neighbor $S^z_iS^z_j$ couplings.
The ground state phase diagram of the spin-chain model is studied by
exact diagonalization of finite chains up to $N=28$ sites, as well as
through analytical techniques.
We find four phases in the phase diagram: two ordered phases in which the
spins have a N\'eel-type of long range order (an unreconstructed and a
missing-row reconstructed phase, in the surface language), a spin liquid
phase (representing a rough surface), and an intermediate dimer phase which
breaks translational invariance and has a doubly degenerate ground state,
corresponding to a disordered flat surface.
The transition from the $(2\times 1)$ reconstructed phase to the disordered
flat phase belongs to the $2D$ Ising universality class.
A critical (preroughening) line with varying exponents separates the
unreconstructed phase from the disordered flat phase.
The possible experimental signatures of the disordered flat phase are
discussed.
\end{abstract}
%
%%%%%%%%%%%%%%%%%%%%%%%%%%%%%%%%%%%%%%%%%%%%%%%%%%%%%%%%%%%%%%%%%%%%%%%%%%
%                                   TEXT
%%%%%%%%%%%%%%%%%%%%%%%%%%%%%%%%%%%%%%%%%%%%%%%%%%%%%%%%%%%%%%%%%%%%%%%%%%
\newpage
\section{Introduction}

Surfaces of fcc metals, in particular (110) faces, display a variety of phase
transitions which have been the subject of considerable experimental and
theoretical work.

Experimentally, (110) faces of some fcc metals - like Au or Pt - are
reconstructed at low temperature into a $(2\times 1)$ missing-row structure,
whereas other metals - like Ag, Ni, Cu, etc.\ - are not.
As the temperature is raised, reconstructed surfaces tend to show two
separate transitions: a deconstruction transition occurs first, followed, at
a higher temperature, by a roughening transition.\cite{Au_exp,Pt_exp}
On unreconstructed surfaces, instead, only a roughening transition has been
well characterized.\cite{Cu_exp,Ni_exp}

On the theoretical side, an interesting and nontrivial interplay has
been anticipated between in-plane ordering, related to reconstruction,
and vertical ordering, related to roughening.\cite{Erio_87}
Since then many studies have been devoted to the problem.
\cite{Vil_Vil_88,KJT,Vil_Vil_91,MdN_92,Bal_Kar,Mazzeo,Bernasconi}

The situation is somewhat different for the two cases, the unreconstructed
and the missing-row reconstructed. For the unreconstructed case, den Nijs
has argued that the phase diagram should be qualitatively similar to that
of a simple cubic (100) surface.\cite{MdN_90} In particular, he proposes that
Ag and Pd (which do not reconstruct) are good candidates for the realization of
{\em preroughening\/}, an additional critical variable-exponent transition
from a low temperature ordered phase to an intermediate temperature
{\em disordered flat\/} phase, previously identified in the context of
restricted solid-on-solid models for simple cubic (100) surfaces.\cite{MdN_Rom}
The unreconstructed surface possesses a N\'eel-type of order parameter,
which characterizes $(1\times 1)$ order
\begin{equation} \label{p1x1:eqn}
P_{(1\times 1)} = \frac{1}{N} \langle \sum_{\bf r} h_{\bf r}
e^{i {\bf G}\cdot {\bf r} } \rangle \;.
\end{equation}
$P_{(1\times 1)}$ can take two opposite values at low temperature and
vanishes at and above preroughening.\cite{Mazzeo,PBW:NOTA}
(Here $h_{\bf r}$ is the height of each surface atom, $N$ is the number
of $(1\times 1)$ cells of the surface, and ${\bf G}=(2\pi/a_x){\bf \hat{x}}$
is the first non-zero reciprocal vector of the lattice, see Fig.\ 3.)

For the reconstructed case, one has {\em four\/} ground states to deal with
(instead to the two ground states of an unreconstructed surface), and these
can be classified, according to den Nijs, \cite{MdN_92} by a clock variable
$\theta=0,\pi/2,\pi,3\pi/2$, see Fig.\ 4.
The most elementary extended defects which one can consider are {\em steps\/},
which change the average height by $\pm 1$ and the reconstruction variable
by $\Delta\theta=\pi/2$ ({\em clockwise\/} or $(3\times 1)$ steps) or
$\Delta\theta=-\pi/2$ ({\em anti-clockwise\/} or $(1\times 1)$ steps), and
{\em Ising wall defects\/} which do not change the average height, and change
the reconstruction variable by $\Delta\theta=\pi$.

The interplay between reconstruction and height degrees of freedom makes
this situation particularly interesting.
The problem has been tackled using two different lines of approach.
The first line of approach consists in identifying heuristically the relevant
defects which play a role into the problem, and building up models which
describe their behaviour.
In this framework, den Nijs has formulated a
{\em four state clock-step model\/} on a length scale somewhat
larger than the microscopic one.\cite{MdN_92}
On each cell of a coarse-grained lattice an integer variable
$h_{\bf r}$ - representing the average height in the cell - as well as a clock
reconstruction variable $\theta_{\bf r}=0,\pi/2,\pi,3\pi/2$ are defined.
By assumption, the only actors appearing in the model are
Ising wall defects - for which $(\Delta\theta =\pi,\Delta h=0)$ -
as well as clockwise, $(\Delta\theta =\pi/2,\Delta h=\pm 1)$, and
anti-clockwise, $(\Delta\theta =-\pi/2,\Delta h=\pm 1)$, steps.
den Nijs found that when both types of steps have the same energy -
the so-called {\em zero chirality limit\/} - the model displays two possible
scenarios: (i) if the parameters are such that the energy of an Ising wall
$E_w$ is less than roughly twice the energy of a step $E_s$, by increasing the
temperature one goes from a reconstructed phase to a disordered flat phase
with an Ising transition, and then to a rough phase with a conventional
Kosterlitz-Thouless (KT) transition; (ii) in the other case - $E_w>2E_s$ - the
system undergoes a single roughening-induced deconstruction
transition.\cite{MdN_92}
Within the same line of approach, but in the so-called strong chirality
limit (only clockwise steps allowed), den Nijs \cite{MdN_92} and,
independently, Balents and Kardar \cite{Bal_Kar} have tackled the problem by
mapping steps into the world-lines of 1D fermions, the Ising wall being
represented by a site doubly occupied by a pair of up and down steps.
The hamiltonian describing such a 1D fermion problem is essentially a
Hubbard model with an extra term describing dislocations, i.e., the merging of
four steps at the boundary of finite terraces on the surface.
Scenario (i) above, realized for an attractive Hubbard $U$, is essentially
recovered in this fermion mapping.
As for scenario (ii), corresponding to $U>0$, the roughening-induced
deconstruction is now substituted by a doublet of successive phase transitions:
first a Pokrovsky-Talapov (PT) transition to a ``floating reconstructed''
rough phase, and then a further KT transition to a rough disordered
phase.\cite{MdN_92,Bal_Kar}

While these approaches are interesting {\em per se\/}, in that they identify
potential scenarios for these surfaces, their outcome is obviously
connected to their initial set of assumptions (i.e., identification of the
relevant defects, and choice of their mutual interactions).

A more direct line of approach consists in working with solid-on-solid
models, which - while still very idealized - are fully microscopic in nature.
Kohanoff {\em et al.\/}, and, more recently, Mazzeo {\em et al.\/}
have introduced a lattice solid-on-solid (SOS) model which is able to deal
with both unreconstructed and reconstructed situations.\cite{KJT,Mazzeo}
Their model is a simple modification of the exactly solvable body-centered
solid-on-solid (BCSOS) model \cite{BCSOS:exact} obtained by adding a further
neighbor interaction, and is quite closely related to the model we will
consider here.
Using a Monte Carlo simulation Mazzeo {\em et al.\/} study two separate
points in its phase diagram, intended to describe the case of reconstructed
Au(110) and that of unreconstructed Ag(110).\cite{Mazzeo}
For the reconstructed case they confirm a two-transition scenario: an
Ising deconstruction transition to a disordered flat phase, followed by a
conventional KT roughening transition.
In the unreconstructed case, quite interestingly, they also find two
transitions: a non-Ising critical one leading to a ``sublattice disordered''
flat phase, followed again by a KT roughening.\cite{Mazzeo}

The model we present in this paper is probably the simplest generalization of
the BCSOS model which is able to describe a missing-row $(2\times 1)$
reconstructed phase as well the unreconstructed one. It essentially consists
of a BCSOS model in which the sign of the, say, x-coupling can have either
sign - the negative sign favoring the occurrence of reconstruction - and a
further neighbor interaction in the x-direction is added in order to stabilize
the reconstructed case against faceting into a $(111)$ surface.
The model is indeed quite similar to the one introduced by Mazzeo
{\em et al.\/}, except for the choice of the non nearest neighbor stabilizing
interaction.
It has the virtue of being easy to map into a one dimensional quantum spin-1/2
model of the Heisenberg type with further neighbor $S^z_iS^z_j$ interactions.
Such a mapping provides a quite powerful way of investigating the full phase
diagram of the model, without having to resort to Monte Carlo simulations.
Moreover, the insight gained by relating our problem to the large body of
knowledge in the field of one-dimensional quantum systems will appear
quite evidently.

One of the motivations of our work was a deeper understanding of the nature
of the disordered flat phase occurring in this kind of simple modifications of
the BCSOS model.
In this respect, our study was greatly inspired by the work of den Nijs
and Rommelse, in particular by their discussion on the many similarities
between the physics of the spin-1 and spin-1/2 chain.\cite{MdN_Rom}
Within our simple model, we explicitly demonstrate that, indeed, a spin-1/2
dimer phase is a natural candidate to representing the disordered flat
phase of an fcc (110) surface.
{}From a surface science viewpoint, a better understanding of the phase
diagram and of the nature of the disordered flat phases of these surfaces
should
help building future experiments directed at their study.

Our results are summarized by the two phase diagrams of Figs.\ 1 and 2,
pertaining to the spin and to the SOS model, respectively.
We find four phases in the phase diagram: two ordered phases in which the
spins have N\'eel-type $(1\times 1)$ and $(2\times 1)$ long range order
(an unreconstructed and a missing-row reconstructed phase, in the surface
language), a spin liquid phase (representing a rough surface), and an
intermediate dimer phase which breaks translational invariance and has a
doubly degenerate ground state (a disordered flat surface).
The transition from the $(2\times 1)$ reconstructed phase to the disordered
flat phase belongs to the $2D$ Ising universality class.
A critical line with varying exponents separates the unreconstructed
phase from the disordered flat phase.
Many of the above mentioned features are, we believe, quite robust and should
hold for more complicated microscopic models as well.
Several scenarios are possible in our phase diagram: First of all, the
two-transition scenario for reconstructed surfaces, i.e., Ising deconstruction
followed by roughening; Next, the possibility - at least in principle - of a
first-order transition from a reconstructed to an unreconstructed surface,
followed by a preroughening transition to a disordered flat phase, and
finally by roughening; Finally, for unreconstructed surfaces,
either a single roughening transition or two transitions (preroughening
followed by roughening), depending on the actual location of the
preroughening line, which is highly model dependent.

The rest of the paper is organized as follows. Section 2 presents the SOS model
and its spin-chain counterpart. In section 3 we describe how the spin-chain
phase diagram is obtained. In section 4, finally, we discuss the physics of
the disordered flat phase in relationship to the dimer phase of the spin
model, and present our conclusions. Relevant technical material on the
spin-chain mapping is contained in the appendix.

\section{A solid on solid model and its spin-chain counterpart}

The model we want to study is defined on a lattice, schematically shown in
Fig.\ 3, which is comprised of two interpenetrating
sublattices, conventionally referred to, hereafter, as the white (W) and
the black (B) sublattice.
In the ideal fcc (110) surface the two sublattices are rectangular, with
$a_x=\sqrt{2} a_y$, and one of them sits above the other a distance
$a_z=a_y/2$.
The notation $(i,j)$ we use for the sites is such that the W and B sublattices
are characterized, respectively, by even and odd values of $i$, whereas
the value of $j$, see Fig.\ 3, is the same for a horizontal
row of W sites and the row of B sites immediately above.
To each site of the lattice we associate a height variable $h_{i,j}$ which
can take only {\em integer\/} values (we fix $a_z=1$).
Moreover, a BCSOS-type of constraint is assumed to hold, i.e., the height
difference between each site and its nearest neighbors (belonging to the
other sublattice) are forced to be $\pm 1$. As a consequence, $h_{i,j}$ are
even, say, on the W sublattice and odd on the B sublattice.
The hamiltonian for the model is written as
\[
{\cal H}={\cal H}_0+{\cal H}_1 \;,
\]
where ${\cal H}_0$ is the BCSOS piece
\begin{equation} \label{mod1:eqn}
{\cal H}_0 = \sum_{i=1}^{N_x} \sum_{j=1}^{N_y}
 [ K_{2y} (h_{i,j}-h_{i,j+1})^2
+ K_{2x} (h_{i+1,j}-h_{i-1,j})^2 ] \;,
\end{equation}
describing next-nearest neighbor interactions with different coupling
strengths in the two directions, and
\begin{equation} \label{mod2:eqn}
{\cal H}_1 = \sum_{i=1}^{N_x} \sum_{j=1}^{N_y}
K_4 \; (h_{i+2,j}-h_{i-2,j})^2 \;,
\end{equation}
(with $K_4\ge 0$) is a fourth neighbor interaction term in the x direction.
Periodic boundary conditions are assumed to hold in both directions.
$K_{2y}$ will be always assumed to be {\em positive\/} and is generally the
largest energy in the problem, since the y direction is the direction of the
missing-rows we are trying to describe. The coupling in the x direction,
$K_{2x}$, can instead have either sign as long as $K_4>0$ whenever
$K_{2x}<0$, for obvious stability reasons. A negative $K_{2x}$ will quite
clearly favour a missing-row type of reconstruction.

The classical $T=0$ ground states for our model are easy to work out,
as a function of the dimensionless ratio ${\cal K}=K_{2x}/K_4$.
One finds that for ${\cal K}>0$ the ground state is an unreconstructed
$(1\times 1)$ surface, whereas for $-4<{\cal K}<0$ it is a $(2\times 1)$
missing-row structure (with missing rows in the y-direction),
see Fig.\ 4.
For $-8<{\cal K}<-4$ a sequence of larger periodicity
- $(3\times 1),\dots,(n\times 1),\dots$ - missing-row phases is obtained,
whereas for ${\cal K}<-8$ the model no longer describes a (110) surface.

When $K_4=0$ and $K_{2x}>0$, we recover the BCSOS model which can be exactly
solved through a mapping to the six-vertex model, \cite{BCSOS:exact} and
shows a {\em single transition\/}, of the Kosterlitz-Thouless (KT) type,
between a low temperature ordered (unreconstructed) flat phase and a high
temperature disordered rough phase.

To proceed further, we map the model in
Eqs.\ (\ref{mod1:eqn}-\ref{mod2:eqn}) into a one-dimensional (1D) quantum
spin-1/2 hamiltonian.
Details about this mapping are given in the appendix. It suffices here to say
that the gist of the method consists in viewing the y-direction as the
(imaginary) time direction of an appropriate 1D quantum problem, whose
hamiltonian $H_S$ is selected in such a way that the matrix elements of
the imaginary-time evolution operator $e^{-\tau H_S}$ coincide, in the
so-called {\em time-continuum limit\/}, with the matrix elements of the
classical transfer matrix of the original problem.\cite{NOTA1}
The crucial physical requirement is that the ``time'' direction must
coincide with the hard direction of the classical problem, i.e., the coupling
in the y direction has to be much stronger than the other couplings.
Anisotropy is not expected to play a role in the phase diagram of the model,
and is, moreover, physically quite appropriate in the present case.
The nearest-neighbor height constraint suggests that the relevant degrees
of freedom can be described by introducing a spin-1/2 Hilbert space,
and mapping the height difference between nearest-neighbors within the
same ``time-slice'' - the dashed line in Fig.\ 3 - to the
z-component of the usual spin operator
\begin{equation}
S^z_i \longleftrightarrow \frac{1}{2} (h_{i+1,0}-h_{i,0})  \;.
\end{equation}
Fig.\ 4 illustrates explicitly the mapping of the $(1\times 1)$
and $(2\times 1)$ ground states in terms of spin configurations along a
time-slice.
In terms of spin operators, the quantum hamiltonian reads
\begin{equation} \label{spin_mod:eqn}
H_S = -\frac{J}{2} \sum_{i=1}^{N_x} [S^+_i S^-_{i-1} + S^-_i S^+_{i-1}]
+ \sum_{i=1}^{N_x} [J_z S^z_i S^z_{i-1}
+ J_2 S^z_i S^z_{i-2}
+ \frac{J_2}{2} S^z_i S^z_{i-3}] \;,
\end{equation}
where the spin couplings are related to the original couplings as follows:
\begin{eqnarray} \label{coup_relat:eqn}
J   &=& 2 \exp{(-4\beta K_{2y})} \nonumber \\
J_z &=& 8 \beta (K_{2x} + 3 K_4) \nonumber \\
J_2 &=& 16 \beta K_4 \;,
\end{eqnarray}
and $\beta=1/k_BT$.
Periodic boundary conditions on the heights are inherited by the spin model,
and also imply that we need to work in the spin sector with zero total
magnetization.
It is well known that the mapping is such that the free energy per site of
the classical problem is related to the ground state energy per site of the
one-dimensional quantum problem, i.e., $\beta f = \epsilon_{GS}$.\cite{Kogut}
The temperature clearly enters through the spin couplings, see Eqs.\
\ref{coup_relat:eqn}, so that any temperature singularity of the classical
free energy can be seen as a ground state energy singularity for the
quantum problem as a function of the couplings $J_z/J$ and $J_2/J$.
Moreover temperature averages for correlation functions of the classical
problem can be likewise rewritten in the form of ground state averages
for the corresponding quantum correlation function.\cite{Kogut}
In summary, to obtain information about the temperature phase diagram of
the classical model we need to study the {\em ground state phase diagram\/}
of the spin-chain model.

As it is, the spin model in Eq.\ \ref{spin_mod:eqn} has not been previously
studied, to our knowledge.
A similar model, however, has been discussed in Ref.\ \cite{Hal_82}, for
instance, and has been the topic of a quite extensive literature.
It differs from our model in that the second neighbor coupling is taken to
be spin isotropic, and the third neighbor coupling is missing, i.e.,
\begin{equation} \label{Hal_mod:eqn}
H = -\frac{J}{2} \sum_{i=1}^{N_x} [S^+_i S^-_{i-1} + S^-_i S^+_{i-1}]
+ \sum_{i=1}^{N_x} [J_z S^z_i S^z_{i-1}
+ J_2 {\vec S}_i \cdot {\vec S}_{i-2}] \;.
\end{equation}
We will see later on that many features of the phase diagram of the model
in Eq.\ \ref{Hal_mod:eqn} are indeed present in the phase diagram of our model
too. The only exception is an extra phase appearing in our model for
sufficiently large $J_2$, a phase which corresponds, in the surface language,
to an ordered $(2\times 1)$ missing-row structure.

\section{Analysis of the quantum spin-chain model}

We start this section by defining the interesting order parameters and
correlation functions that we need to look at in order to determine the
phase diagram for our model.
A quite direct way of defining an order parameter for the quantum spin-chain
case, is to consider the ``square'' of the appropriate
``staggered magnetization'' $N^{-1} \sum_j e^{i(\pi/p)j} S^z_j$.
The result can be trivially cast in terms of the Fourier transform
$S^{zz}(k)$ of the spin-spin correlation function (the ``structure factor''):
\begin{equation}
P^2_{p\times 1} = \lim_{N\rightarrow \infty} \frac{1}{N} \sum_j
e^{i(\pi/p)j} \langle S^z_0 S^z_j \rangle =
\lim_{N\rightarrow \infty} \frac{1}{N} S^{zz}_N(\pi/p) \;,
\end{equation}
where the average has to be intended as a ground state average.
Clearly, ordered phases will be signalled by a structure factor $S^{zz}_N(k)$
which {\em diverges linearly\/} with $N$ - for the appropriate k-vector - as
the length of the system $N\rightarrow \infty$.
More specifically, $S^{zz}_N(\pi)$ will diverge for a phase with
$\uparrow\downarrow\uparrow\downarrow$ (N\'eel) long range order (LRO),
a $(1\times 1)$ unreconstructed phase in the surface language, see Fig.\ 4.
A divergence of $S^{zz}_N(\pi/2)$ will instead indicate
$\uparrow\uparrow\downarrow\downarrow$ LRO, i.e., a missing-row $(2\times 1)$
reconstructed surface, see Fig.\ 4.

Another quantity one needs to look at, in a SOS model, is the height-height
correlation function
\begin{equation}
G({\bf r}-{\bf r'}) = \langle [h_{\bf r} - h_{\bf r'}]^2 \rangle \;,
\end{equation}
which distinguishes a {\em flat\/} [$G(r)<\mbox{const}$ as
$r\rightarrow\infty$]
from a {\em rough\/} [$G(r)\approx\ln{(r)}$ as $r\rightarrow\infty$] phase.
The height-height correlation function has a simple translation in the
spin-chain model. Restricting our consideration to correlations within the same
horizontal row, say j=0, we have
\begin{equation}
G(n) = \langle [h_{n,0} - h_{0,0}]^2 \rangle =
4\sum_{i,j=0}^{n-1} \langle GS| S^z_i S^z_j |GS\rangle =
n + 8\sum_{i=1}^{n} (n-i) \langle GS| S^z_0 S^z_i |GS\rangle \;,
\end{equation}
where the last equality follows from assuming translational invariance for
the spin-spin correlation function. Notice, in passing, that the term linear
in $n$ is exactly cancelled by working at zero total magnetization since, in
that case,
\[ \sum_{i\ne 0} \langle GS| S^z_0 S^z_i |GS\rangle =
-\langle GS| S^z_0 S^z_0 |GS\rangle = -\frac{1}{4} \;. \]
Moreover, it is easily checked that whenever the spin-spin correlation
function $\langle S^z_0 S^z_n \rangle$ possesses a large distance uniform term
of the type $-K/(2\pi^2n^2)$, $G(n)$ will diverge {\em logarithmically\/} as
\begin{equation} \label{G_div:eqn}
G(n) = \frac{4K}{\pi^2} \log{(n)} + \cdots \hspace{20mm}
n \rightarrow \infty \;,
\end{equation}
signalling a rough phase.

We now discuss the information that one can get on different regions of the
phase diagram from analytical approaches.
In the limit $J_z/J\rightarrow \infty$ and $J_2/J\rightarrow \infty$
things are very easy to work out.
Setting $J=0$, one finds that for $J_2<(2/3)J_z$ the N\'eel states
$|\uparrow\downarrow\uparrow\downarrow \cdots \rangle$ and
$|\downarrow\uparrow\downarrow\uparrow \cdots \rangle$ are the two possible
ground states, whereas for $J_2>(2/3)J_z$ there are four possible ground states
\begin{eqnarray}
|1\rangle &=& |\uparrow\uparrow\downarrow\downarrow\uparrow\uparrow
\downarrow\downarrow  \cdots \rangle \nonumber \\
|2\rangle &=& |\downarrow\uparrow\uparrow\downarrow\downarrow\uparrow
\uparrow\downarrow  \cdots \rangle \nonumber \\
|3\rangle &=& |\downarrow\downarrow\uparrow\uparrow\downarrow\downarrow
\uparrow\uparrow  \cdots \rangle \nonumber \\
|4\rangle &=& |\uparrow\downarrow\downarrow\uparrow\uparrow\downarrow
\downarrow\uparrow  \cdots \rangle \;.
\end{eqnarray}
Such a fourfold degeneracy of the ground state reflects the fourfold degeneracy
of the $(2\times 1)$ missing-row surface, as illustrated in Fig.\ 4.

We now turn to the region of the phase diagram close to $J_z=J_2=0$.
Clearly, when $J_2=0$ (i.e., $K_4=0$) the problem reduces to the XXZ Heisenberg
chain - the quantum spin-chain counterpart of the BCSOS model - whose physics
is well known. For $|J_z/J|\le 1$ the model is gapless and critical, i.e.,
the ground-state spin-spin correlation functions decay algebraically at
large distances \cite{Hal_82,Sch_Zim}
\begin{eqnarray} \label{XXZ_corr_exp:eqn}
\langle S^z_0 S^z_n \rangle &\approx &  \cos{(\pi n)}
\frac{A}{n^{\eta_z}} -\frac{K}{2\pi^2n^2} \,+\, \cdots \nonumber \\
\langle S^+_0 S^-_n \rangle &\approx & \cos{(\pi n)} \frac{B}{n^{\eta}} \,+\,
\cdots \;.
\end{eqnarray}
The exponents $\eta$ and $\eta_z$ are exactly known:
\begin{equation}
\eta = \frac{1}{\eta_z} = \frac{1}{\pi} \arccos{(-J_z/J)} \;.
\end{equation}
{}From the previous discussion - see Eq.\ \ref{G_div:eqn} - it follows
immediately that such a phase represents a {\em rough surface\/}.
For $J_z/J>1$ the model has a gap and a N\'eel order parameter, and
represents the already mentioned unreconstructed $(1\times 1)$ phase (the
BCSOS flat phase).
The transition at $J_z/J=1$ is of the KT-type (the BCSOS roughening
transition).
Consider now the case $J_2>0$.
By applying a Wigner-Jordan transformation we can equivalently rewrite
our spin model in terms of a spinless fermion model
\begin{equation} \label{fermion_mod:eqn}
H_F = -J \sum_{k}^{BZ} \cos{(k)} \; c^+_k c_k +
+ \frac{1}{N} \sum_{q}^{BZ} V(q) \rho(q) \rho(-q) \;,
\end{equation}
where $\rho(q)=\sum_k c^+_{k} c_{k+q}$ is the density operator and
$V(q)=J_z\cos{(q)}+J_2\cos{(2q)}+(J_2/2)\cos{(3q)}$ is the interaction
potential.\cite{NOTA2}
Zero total magnetization for the spin model implies a half-filled band for the
fermion model.
Applying standard techniques in the theory of 1D fermionic systems,
i.e., linearizing the band around the two Fermi points at $k_F=\pm\pi/2$ and
keeping only interaction processes around the Fermi points, we can write down
an effective continuum field theory (``g-ology model'').\cite{Sol}
Let $a_{\pm,k}$ be the destruction operator for right-moving ($+$) and
left-moving ($-$) fermions around the two Fermi points, and let
$\rho_{\pm}(q)$ be the corresponding density operators.
The g-ology model reads:
\begin{eqnarray}
H_{\rm eff} &=& v_F\sum_{k} k[a^+_{+,k}a_{+,k} - a^+_{-,k}a_{-,k}] \nonumber \\
&& +\,\frac{g_4}{L}\sum_q[\rho_+(q)\rho_+(-q)+\rho_-(q)\rho_-(-q)]
 \,+\,\frac{g_2}{L}\sum_q[\rho_+(q)\rho_-(-q)] \nonumber \\
&& +\,\frac{g_3}{L}\sum_q[\rho^{(+)}(q)\rho^{(+)}(-q) +
\rho^{(-)}(q)\rho^{(-)}(-q)] \nonumber \\
\end{eqnarray}
where $\rho^{(+)}(q)=\sum_k a^+_{+,k-q} a_{-,k}$ and
$\rho^{(-)}(q)=\sum_k a^+_{-,k-q} a_{+,k}$. To lowest order, the couplings
turn out to be $g_4=J_z+(3/2)J_2$, $g_2=4J_z+2J_2$ and $g_3=-2J_z+J_2$.
The $g_3$ term describes {\em umklapp\/} processes.
In absence of $g_3$, the $g_4$ and $g_2$ terms define a spinless fermion
Luttinger model which can be exactly solved (at least when the band cut-off
is removed and one works with infinite linear bands).
Notice that $J_2$ competes with $J_z$ in determining the sign of the
umklapp term coupling constant $g_3$.
This is a quite general condition under which a preroughening line, separating
two different gapped phases, is obtained.
The continuum model coincides, apart from slightly different weak coupling
values for $g_2$,$g_4$, and $g_3$, with the corresponding continuum model
derived for the hamiltonian with spin isotropic second neighbor interaction
in Eq.\ \ref{Hal_mod:eqn}. \cite{Hal_82}
The physics of the two models should indeed be the same in the region close
to the origin, and we can therefore borrow most of the conclusions of
Ref.\ \cite{Hal_82}.
Summarizing the results of Ref.\ \cite{Hal_82}, we can say that
there is a whole region of the phase diagram - for sufficiently small $J_2$
and $J_z$ - in which the model is within the basin of attraction of a spinless
Luttinger model, i.e., $g_3$ scales to zero upon renormalization.
In short, borrowing Haldane's terminology, \cite{Hal_81} we have a
{\em Luttinger liquid\/}.
As a result, the spin-spin correlation functions will have exactly the same
large distance behaviour as in Eq.\ \ref{XXZ_corr_exp:eqn}, with suitable
coupling dependent exponents $\eta_z$ and $\eta$.
Moreover, quite generally, $\eta_z=1/\eta$ and the constant $K$ is related
to the exponents, $K=\eta_z/2$.\cite{Hal_81}
In the presence of the umklapp term ($g_3\ne 0$), such a Luttinger liquid
becomes {\em unstable\/} towards phases having a gap in the excitation
spectrum whenever the constant $K$ goes to $1/2$
(or $\eta_z\rightarrow 1$).\cite{Sol}
Such a transition is predicted to be in the KT universality class.
There is nevertheless a whole line of points, inside the gapped phases, on
which effectively the coefficient of the umklapp term $g_3$ is exactly zero.
Such a line will consist of critical points in which the model still
behaves as a Luttinger liquid with a varying exponent $\eta_z<1$.
The theory of Ref.\ \cite{Hal_82} predicts also the nature of the two
gapped phases to which the Luttinger liquid becomes unstable on either side of
the ``$g_3=0$'' line.
On one side, the already mentioned $\uparrow\downarrow\uparrow\downarrow$
N\'eel phase is found, whereas on the other side a
{\em spontaneously dimerized phase\/} should occur.\cite{Hal_82}

To determine the actual location of the different lines in the phase diagram
we need to resort to a numerical study. We have performed a finite size
scaling analysis of exact diagonalization data for chains up to $N=28$ sites.
The location of the KT lines terminating the Luttinger liquid (rough) phase
are determined from the size scaling of the gaps as done, for instance, in
Ref.\ \cite{Sch_Zim}.
More in detail, the finite size gaps from the ground state to the first
excited state in the spin-1 and spin-0 sectors will determine $\eta$ and
$\eta_z$, respectively,
\begin{eqnarray} \label{gap_scal:eqn}
N [E_N(S\!=\!1,k\!=\!0) - E_N(S\!=\!0,k\!=\!0)] &=&
    \pi v_s \eta + \cdots \nonumber \\
N [E_N(S\!=\!0,k\!=\!\pi) - E_N(S\!=\!0,k\!=\!0)] &=&
    \pi v_s \eta_z + \cdots \;,
\end{eqnarray}
as $N\rightarrow \infty$.
The sound velocity $v_s$ can be similarly determined by knowing the gap from
the ground state to the first excited state of smallest momentum
\begin{equation}
N [E_N(S\!=\!0,k\!=\!2\pi/N) - E_N(S\!=\!0,k\!=\!0)] = 2 \pi v_s + \cdots \;.
\end{equation}
An independent way of estimating $\eta_z$ (or $K$) comes from
the $q\rightarrow 0$ limit of the structure factor $S^{zz}_N(q)$.
In the Luttinger liquid phase we have, for $q\rightarrow 0$,
$S^{zz}(q)=(K/2\pi)q + o(q)$, and therefore
\begin{equation} \label{k_sq:eqn}
\lim_{N\rightarrow \infty} N S^{zz}_N(q=2\pi/N) = K \;.
\end{equation}
(The advantage of the latter relationship is that the sound velocity $v_s$
does not appear at all.)
As an illustration, Fig.\ 5 shows the finite-size results
obtained for $v_s \eta$ (a) and $v_s$ (b) along the line at $J_z/J=1$ for
several values of $J_2/J$.
Power law extrapolations of data for $v_s\eta$, $v_s\eta_z$, and $v_s$ yield
the results for the exponents summarized in Fig.\ 6.
Quite unambiguously, a transition - at which $\eta_z=1/\eta=1$ - occurs
for $J_2/J\approx 1.2\div 1.3$.
An independent confirmation of such a result comes from the structure
factor data shown in Fig.\ 7 (see Eq.\ \ref{k_sq:eqn}), from
which it is quite clear, for instance, that $J_2/J=1.4$ is already beyond
the transition point (the dashed line in Fig.\ 7, at $K=\eta_z/2=1/2$).
A similar procedure has allowed us to determine the location of the KT lines
in the phase diagram of Fig.\ 1.

The location of conventional second order transitions (which are expected to
terminate the large $J_2$ ordered phase) are determined from the ``specific
heat'' $C_v=\partial^2\epsilon_{GS}/\partial J_2^2$. Fig.\ 8 shows
$C_v$ for chains up to $N=28$ sites as a function of $J_2$ at $J_z/J=1$.
The location of the maximum of $C_v$ quickly saturates to our estimate of the
critical value of $J_2$, $J_2^{(c)}\approx 1.66J$. Plotting $C_v$ at the
estimated $J_2^{(c)}$ versus the logarithm of the system size (see inset of
Fig.\ 8), we conclude that $C_v$ diverges logarithmically, i.e.,
the critical exponent $\alpha=0$.
In order to calculate another exponent, we consider the structure factor
$S^{zz}_N(\pi/2)$.
In the ordered $(2\times 1)$ phase $S^{zz}_N(\pi/2)$ diverges linearly with
$N$. Exactly at the critical point, however, the spin-spin correlation
function will decay as a power law at large distances,
$\langle S^z_0 S^z_n \rangle \approx \cos{(n\pi/2)}/n^{\eta'}$.
If the power law exponent $\eta'$ happens to be strictly less than $1$, the
structure factor will diverge at the critical point as
\begin{equation}
S^{zz}_N(\pi/2) \approx (\mbox{const}) \; N^{1-\eta'} \hspace{15mm}
\mbox{at $J_2^{(c)}$ as $N\rightarrow \infty$.}
\end{equation}
Fig.\ 9 shows a log-log plot of $S^{zz}_N(\pi/2)$ versus
$N$ for $J_z=J$ and different values of $J_2$ around $J_2^{(c)}\approx 1.66J$.
A good straight line fit is obtained only for $J_2=1.66J$, and the slope of
the straight line is $\approx 0.75$ (the solid line in Fig.\ 9).
This implies $\eta'=1/4$.
The transition belongs, therefore, quite unambiguously to the $2D$ Ising
universality class.

As far as the predicted dimer phase is concerned, there are clear numerical
indications that we are indeed dealing with a doubly degenerate ground state
with a gap above it. Fig.\ 10 shows, for instance, the
size dependence of the gap to the first excited $S=1$ state for $J_2/J=1.6$
and $J_z/J=1$.
A pronounced increase is seen when one plots $N\Delta E_N$ versus
$1/N$, a quite clear hint that $\Delta E_N$ is saturating to a constant.
On the contrary, the gap to the first excited $S=0$ state of momentum $\pi$ is
closing up, as shown in the inset of Fig.\ 10, hinting that
such a state is actually degenerate with the ground state in the infinite
volume limit. Notice that, with chains of length $N$ up to $28$, a reasonably
clear gap, as the one illustrated in Fig.\ 10 for the $S=1$ state,
is seen only quite far from the transition to the dimer phase.
The same argument applies to a direct test of LRO.
We have in fact measured a dimer-dimer correlation function
\begin{equation}
D_{ij} = \langle (S^z_{i-1} S^z_i) (S^z_{j-1} S^z_j) \rangle \;,
\end{equation}
which is predicted to acquire LRO, $D_{ij}\approx C (-1)^{i-j}$, in the dimer
phase.
Once again, although a clear tendency to LRO is seen within our ``small
chains'', a full fledged signal of LRO, i.e., the $k=\pi$ structure factor of
$D_{ij}$ diverging linearly with $N$, is almost never seen, except very far
from the transition. All these features point into the direction of small
gaps and large correlation lengths, which are numerically difficult to assess.

The location of the variable exponent ($g_3=0$) line in between the N\'eel
and the dimer phase - characterized by $1/4<\eta_z<1$ \cite{Hal_82} - has
been obtained from a combination of the gap scaling procedure in
Eq.\ \ref{gap_scal:eqn}, and from the power law divergence of the structure
factor $S^{zz}_N(\pi)$ (using the procedure discussed above for assessing the
Ising nature of the second order transition).
The variable exponent line, labelled PM in Figs.\ 1 and 2, can be identified
with the {\em preroughening\/} line introduced by den Nijs and Rommelse in
the context of restricted SOS models for simple cubic (100)
surfaces.\cite{MdN_Rom}
The preroughening line merges with the second order (Ising) line at a
tricritical point M, beyond which a first order line should follow. Such
a first order line should asymptotically tend to approach $J_2=(2/3)J_z$,
the value determined from the $J=0$ limit.

The final phase diagram for the spin model is depicted in Fig.\ 1. Most of the
lines are quantitative although some points are not determined
with great accuracy. The region around the point labelled P shows pronounced
size effects which make the exact location of the lines somewhat more uncertain
than elsewhere. Even more pronounced size effects are seen is the
$J_z\approx 0$ region of the phase diagram where the KT line ending the
spin-fluid (Luttinger) phase approaches the Ising deconstruction line.
Such large size effects, if on the one hand prevent us from giving a clear
answer to the behaviour of the model in that region, are, on the other hand,
a quite probable hint that the Ising and KT lines might merge at some point.
We are at present unable to say if the scenario of a single
``roughening-induced deconstruction'' is indeed at play in our model, and
even less able to say if, in such a case, the Ising and KT nature of the
original lines would simply be superimposed, as seen in Ref.\ \cite{MdN_92}.

A translation of the spin-chain phase diagram in Fig.\ 1, using the coupling
relationships in Eq.\ \ref{coup_relat:eqn}, allows us to determine the
temperature phase diagram for the initial classical SOS model, which is shown
in Fig.\ 2 for $K_4/K_{2y}=0.1$.
The different lines can be obtained, in principle, in a quantitative way from
Fig.\ 1, with discrepancies from the expected exact results which are probably
around few percents, even for not so strongly anisotropic couplings.

\section{The dimer phase as a disordered flat phase}

Let us summarize some well known properties of the dimer phase appearing in our
phase diagram in Fig.\ 1, trying to give an interpretation of them in the
surface language. (See also Ref.\ \cite{MdN_Rom}.)
The dimer phase has a doubly degenerate ground state with a gap above it.
Translational invariance is spontaneously broken. All two-spin correlation
functions decay exponentially to zero at large distances, which implies - see
discussion in section 2 - that the surface is disordered but flat.
There is, however, LRO if one considers four-spin correlation function of
the type $\langle ({\vec S}_{i-1}\cdot {\vec S}_i)
({\vec S}_{j-1}\cdot {\vec S}_j)\rangle$.
To be more specific, let us consider a representative dimer phase point
which corresponds to the exactly solvable case of the model in
Eq.\ \ref{Hal_mod:eqn}, i.e., $J_z=J$ and $J_2=J/2$.
This is a particular case of a large class of spin-chain models with competing
interactions whose ground state is known to be a simple product of singlets
(Madjumar-Gosh models).\cite{Caspers}
Explicitly, for any finite (even) size $N$ there are two ground states
(which go into one another by translation of a lattice spacing)
\begin{eqnarray}
|\Psi_1\rangle &=& |1 2\rangle |3 4\rangle \cdots |N-1 N\rangle \nonumber \\
|\Psi_2\rangle &=& |2 3\rangle |4 5\rangle \cdots |N-2 N-1\rangle
|N 1\rangle \;,
\end{eqnarray}
where $|i j\rangle=|\uparrow\downarrow-\downarrow\uparrow\rangle/\sqrt{2}$
denotes a singlet between sites $i$ and $j$. Expanding the product of singlets
it is easy to check that each time a pair $\uparrow\uparrow$ appears, it must
be followed, sooner or later, by a $\downarrow\downarrow$ pair. The surface
is therefore {\em flat} since a step up is followed necessarily by a step down.

Let us now ask ourselves the following question. Suppose we want to count,
in the surface terminology, the difference in the number of white and black
local maxima in the surface. For simplicity, we restrict our considerations
to sites which are local maxima when considered in the x-direction only.
In the spin language, a local ``maximum'' at site $j$ occurs whenever the
site $j-1$ has spin $\uparrow$ and the site $j$ has spin $\downarrow$.
An operator which ``counts'' the maximum at $j$ is therefore
$(S^z_{j-1}+1/2)(1/2-S^z_j)$.
The difference between white (even $j$) and black (odd $j$) maxima is therefore
measured by the order parameter
$P_{BW}=(2/N)\sum_j e^{i\pi j} (S^z_{j-1}+1/2)(1/2-S^z_j)$.
$P_{BW}$ is odd under translation.
Its value is $1$ on the N\'eel state
$|\uparrow\downarrow\uparrow\downarrow\cdots\rangle$, and $-1$ on the other
N\'eel state $|\downarrow\uparrow\downarrow\uparrow\cdots\rangle$.
(Quite generally, it is different from zero in the whole N\'eel phase of
our phase diagram in Fig.\ 1.)
Consider now the value of $P_{BW}$ on the dimer state $|\Psi_1\rangle$.
Using the elementary results that
$\langle\Psi_1| S^z_{2j-1} S^z_{2j} |\Psi_1\rangle=-1/4$ and
$\langle\Psi_1| S^z_{2j} S^z_{2j+1} |\Psi_1\rangle=0$, and observing that
the terms linear in $S^z$ vanish identically, we arrive at
\begin{equation} \label{PBW_dimer:eqn}
\langle\Psi_1| P_{BW} |\Psi_1\rangle = -\frac{2}{N} \sum_j e^{i\pi j}
\langle\Psi_1| S^z_{j-1} S^z_{j} |\Psi_1\rangle = \frac{1}{4} \;.
\end{equation}
(Similarly, $\langle\Psi_2| P_{BW} |\Psi_2\rangle = -1/4$.)
We clearly recognize, in Eq.\ \ref{PBW_dimer:eqn}, the $zz$-component of the
usual dimer order parameter. Therefore, a surface interpretation of the
dimer order parameter and of the spontaneous breaking of translational symmetry
is that either the white or the black sublattice tend to dominate in the
surface local maxima. This property has been directly verified by a Monte Carlo
simulation of the classical SOS model.\cite{Michele}
More precisely, we have considered an operator $O_{i,j}$ whose value is
$1$ if the site is an actual 2D local maximum, and zero otherwise, and then
constructed the following order parameter \cite{Michele}
\begin{equation}
P^{\rm (surf)}_{BW} = \frac{2}{N_x N_y} \langle \sum_{i,j} (-1)^i O_{i,j}
\rangle \;.
\end{equation}
Our Monte Carlo simulation confirms that $P^{\rm (surf)}_{BW}\ne 0$
(in the thermodynamical limit) in the disordered flat phase of Fig.\ 2.

\section{Order parameters and surface scattering experiments}

As exemplified in the previous section, a disordered flat phase corresponding
to the spin dimer phase should be characterized by a spontaneous breaking of
the translational invariance, and, consequently, by a predominance of one
of the two sublattices in the topmost layer.

These features should be of relevance in the context of surface
scattering experiments.
We discuss here the case of He scattering. The discussion of the X-ray
scattering case can be carried along similar lines, and is dealt with in
Ref.\ \cite{Michele}.
In the kinematical approximation, and within a SOS framework, the coherent
part of the specular peak intensity with perpendicular momentum transfer
in the so-called anti-phase configuration is proportional to
\begin{equation}
I^{coh}({\bf Q}=0,q_z=\pi/a_z) \,\propto\,
|\langle \sum_{i,j} e^{i\pi h_{i,j}} \alpha_{i,j} \rangle |^2 \;,
\end{equation}
where $\alpha_{i,j}$ is an appropriate shadowing factor which takes into
account the physical requirement that peaks scatter more then
valleys.\cite{Mazzeo,Levi}
For our BCSOS-type of model, in which even $i$'s (W sublattice) are associated
to even heights $h_{i,j}$ and odd $i$'s (B sublattice) to odd $h_{i,j}$, one
immediately concludes that $e^{i\pi h_{i,j}}=(-1)^i$ for any allowed
height configuration.
$I^{coh}({\bf Q}=0,q_z=\pi/a_z)$ would therefore be identically zero if all
the atoms were to scatter in the same way ($\alpha_{i,j}=1$ for all $(i,j)$).
In the assumption that only the local peaks scatter efficiently
($\alpha_{i,j}=1$ if $(i,j)$ is a local peak, $\alpha_{i,j}=0$ otherwise)
we obtain that $I^{coh}({\bf Q}=0,q_z=\pi/a_z)$ is proportional to the
square of the order parameter introduced above, \cite{Bernasconi,Michele}
\begin{equation}
I^{coh}({\bf Q}=0,q_z=\pi/a_z) \,\propto\, |P^{\rm (surf)}_{BW}|^2 \;.
\end{equation}
Quite generally, for a reasonably large class of choices of
shadowing factors $\alpha_{i,j}$, the breaking of translational invariance
should guarantee that $I^{coh}({\bf Q}=0,q_z=\pi/a_z)$ is different
from zero (albeit possibly small) in the disordered flat phase considered here.
More precisely, this is so for all the shadowing factors which can be written
in terms of local operators (in the $h_{i,j}$ variables) whose correlation
function is long-ranged in the disordered flat phase.
We mention here a particularly simple choice of shadowing factors, proposed in
Ref.\ \cite{Levi}, which {\em does not\/} involve long-ranged operators:
$\alpha_{i,j}=2-n_{i,j}/2$, where $n_{i,j}$ is the number of
neighbors of the atom in $(i,j)$ which are found at a level higher than the
atom itself. This expression {\em linearly\/} interpolates between
$\alpha=2$ (local maximum) and $\alpha=0$ (local minimum). It can be recast
in the form $\alpha_{i,j}=1-(1/4)\sum_{n.n.} \Delta h_{i,j}$, where
$-\Delta h_{i,j}$ is the height difference between site $(i,j)$ and any of
the four neighboring sites.
Indeed, by exploiting this linearity, it is very simple to show that, such a
choice of $\alpha_{ij}$ leads to a $I^{coh}({\bf Q}=0,q_z=\pi/a_z)$ which is
proportional to the square of the $(1\times 1)$ order parameter
$P_{(1\times 1)}$ (see Eq.\ \ref{p1x1:eqn}),
\begin{equation}
I^{coh}({\bf Q}=0,q_z=\pi/a_z) \,\propto\,
|\langle \sum_{i,j} (-1)^i h_{i,j} \rangle |^2 \;,
\end{equation}
and therefore vanishes at and beyond the preroughening line.

In conclusion we have proposed a simple microscopic SOS model for fcc (110)
surfaces which can be easily mapped into a spin-1/2 quantum chain of the
Heisenberg type with competing $S^z_iS^z_j$ interactions.
The resulting phase diagram is quite simple and clear. Apart from the obvious
unreconstructed and missing-row reconstructed ordered phases, and the high
temperature rough phase, we find a disordered flat phase which shows the
physics of the dimer spin phase. The possible experimental relevance of such a
state has been discussed.
%It breaks translational invariance and one should be able to distinguish which
%sublattice prevails in the top layer.

\section*{Appendix: Derivation of the spin model}

The general idea behind the mapping of a classical statistical mechanics
problem
in $D$ dimension into a quantum problem in $D-1$ (space) dimensions is quite
old, and needs not to be repeated here.\cite{Kogut}
For the problem at hand, the details of the mapping are quite similar to
the procedure sketched in Ref.\ \cite{MdN_Rom} for the spin-1 case.
We briefly report them here for the reader's convenience.

The starting point for the mapping to a spin model is a T-matrix formulation
for the classical partition function
\begin{equation}
{\cal Z} = \sum_{h_{i,j}} e^{-\beta{\cal H}} =
\sum_{h^{(1)}} \cdots \sum_{h^{(N_y)}}
\langle h^{(1)}| \hat{T} |h^{(N_y)}\rangle \cdots
\langle h^{(3)}| \hat{T} |h^{(2)}\rangle \;,
\langle h^{(2)}| \hat{T} |h^{(1)}\rangle \;,
\end{equation}
where $|h^{(j)}\rangle =\{h_{i,j}:i=1,\cdots,N_x\}$ is the jth {\em row
configuration\/} (the dashed line in Fig.\ 3), and
$\hat{T}$ is the classical transfer matrix.
Periodic boundary conditions have been assumed.
It is implicitly assumed that the configurations included in the sum have
to respect the BCSOS restriction constraint.
Indeed, by exploiting this feature, which implies that
$h_{i+1,j}-h_{i,j}=\pm 1$, we can associate to any row configuration
$|h^{(j)}\rangle$ a state $|j\rangle=|S^z_1,S^z_2,\cdots,S^z_{N_x}\rangle$ in
the Hilbert space of a quantum spin-1/2 chain (of length $N_x$) by the
relationship
\begin{equation}
S^z_i \longleftrightarrow \frac{1}{2} (h_{i+1,j}-h_{i,j})  \;.
\end{equation}
(In so doing we actually loose information on the absolute height of the
surface.)
The idea is now to try to reproduce the Boltzmann factors appearing in the
matrix elements of the classical transfer matrix
$\langle h^{(j+1)}| \hat{T} |h^{(j)}\rangle$ by a suitable quantum
operator $T_Q$ in the spin Hilbert space, i.e.\
\begin{equation}
\langle h^{(j+1)}| \hat{T} |h^{(j)}\rangle =
\langle j+1| T_Q |j\rangle \;,
\end{equation}
where $|j\rangle$ and $|j+1\rangle$ are the spin states corresponding to
$|h^{(j)}\rangle$ and $|h^{(j+1)}\rangle$ respectively.
For the hamiltonian in Eqs.\ (\ref{mod1:eqn}-\ref{mod2:eqn}), the
$\hat{T}$-matrix element reads:
\begin{eqnarray}
\langle h^{(j+1)}| \hat{T} |h^{(j)}\rangle \,=\, &&
\exp{ \{ -\sum_{i=1}^{N_x} [ K_{2x} (h^{(j)}_{i+1}-h^{(j)}_{i-1})^2 +
K_{4} (h^{(j)}_{i+2}-h^{(j)}_{i-2})^2 ] \} } \; \times
\nonumber \\
&& \exp{ \{ -K_{2y}\sum_{i=1}^{N_x} (h^{(j+1)}_{i}-h^{(j)}_{i})^2 \} }  \;.
\end{eqnarray}
The first exponential contains interaction terms within row $j$ and it is quite
simple to rewrite it in terms of exponentials of $S^z$ operators only
(see below).
On the contrary, the second exponential contains the interaction which
couples row $j$ to row $j+1$ and is explicitly off-diagonal.
It is not hard to show that, given all the height contraints, the spin
operator $(S^+_i S^-_{i-1} + S^-_i S^+_{i-1})$ does exactly the job of
transforming the spin configuration around site $i$ in row $j$, with
$h_{i}^{(j)}$, into the correct spin configuration of the corresponding site
in row $j+1$ with $h_{i}^{(j+1)}=h_{i}^{(j)}\pm 2$.
Such an operator must therefore be multiplied by $e^{-4K_{2y}}$ if we want
to reproduce the correct Boltzmann weight.
The (diagonal) situation in which $h_{i}^{(j+1)}=h_{i}^{(j)}$ is simply
given by the identity operator in the spin space.
The only tricky point is that $(S^+_i S^-_{i-1} + S^-_i S^+_{i-1})$
{\em does not commute\/} with the pieces involving $S^z$, so that care must
be taken in ordering the various terms.
One can show that the operator $T_Q$ which does the job correctly is given by
\begin{eqnarray}
T_Q &=& \cdots \; T(n+3) T(n) \; \cdots \; T(7) T(4) T(1) \nonumber \\
T(n) &=& [1 + e^{-4\beta K_{2y}} (S^+_n S^-_{n-1} + S^-_n S^+_{n-1})] \times
e^{-8\beta K_{2x}(S^z_n S^z_{n-1}+1/4)} \nonumber \\
&& \hspace{10mm}
\times e^{-4\beta K_4 (S^z_{n-2} + S^z_{n-1} + S^z_{n} + S^z_{n+1})^2} \;,
\end{eqnarray}
where every $T(n)$ for $n=1,\cdots,N_x$ is included in the product which is
however organized in a cyclic way (i.e., with $n$ increasing
by $3$ from right to left and taken modulo $N_x$).
This is not yet a very useful relationship, because, as mentioned above,
the quantum operators appearing in the above expression do not commute.
Our aim is indeed to work in a limit in which all the different factors can
be grouped together under a single exponential of the form $e^{-H_S}$.
This is the case in the so-called {\em time continuum limit\/}, defined by
the fact that $e^{-4\beta K_{2y}}=J/2$ is small,
in which case $1+(J/2)(\cdots)\approx e^{(J/2)(\cdots)}$,
and $|\beta K_{2x}|$ as well as $|\beta K_{4}|$ are also small, so that one
can neglect the commutators coming about in regrouping all the exponentials.
The final result can be cast in the form
\begin{equation}
\hat{T} \longleftrightarrow T_Q \approx e^{-H_S} \hspace{20mm}
\mbox{(Time Continuum Limit)}\;,
\end{equation}
where the spin hamiltonian $H_S$ is given in Eq.\ \ref{spin_mod:eqn}.
The time continuum limit is essentially a limit of high anisotropy.
Whenever anisotropy is not ``relevant'', one obtains a temperature phase
diagram for the classical problem from studying the ground state phase
diagram of the quantum problem.
What is more surprising is that the temperature phase diagram turns out to
be quantitatively accurate (within a few percent) even for situations in which
the classical system is not so strongly anisotropic. This can be explicitly
checked in the 2D Ising case, \cite{Kogut} and in the BCSOS case. \vspace{2cm}

\noindent
ACKNOWLEDGEMENTS - We are warmly grateful to Alberto Parola, Sandro Sorella,
Marco Bernasconi, and Erio Tosatti for many enlightening discussions.
This research was supported by the Italian Ministry of University and
Scientific Research, the ``Istituto Nazionale di Fisica della Materia'',
and the ``Consiglio Nazionale delle Ricerche'' under the
``Progetto Finalizzato `Sistemi Informatici e Calcolo Parallelo' ''.

%%%%%%%%%%%%%%%%%%%%%%%%%%%%%%%%%%%%%%%%%%%%%%%%%%%%%%%%%%%%%%%%%%%%%%%%%
%                               BIBLIOGRAPHY
%%%%%%%%%%%%%%%%%%%%%%%%%%%%%%%%%%%%%%%%%%%%%%%%%%%%%%%%%%%%%%%%%%%%%%%%%

%%%%%%%%%%%%%%%%%%%%%%%%%%%%%%%%%%%%%%%%%%%%%%%%%%%%%%%%%%%%%%%%%%%%%%%%%%
%                                CAPTIONS
%%%%%%%%%%%%%%%%%%%%%%%%%%%%%%%%%%%%%%%%%%%%%%%%%%%%%%%%%%%%%%%%%%%%%%%%%%
\newpage
\begin{center}
{\bf Figure Captions}
\end{center} \vspace{15mm}

\begin{description}

\item[Figure 1]
Ground state phase diagram for the spin-chain model.
PM is a variable exponent (preroughening) line with $1>\eta_z>1/4$. Beyond
the tricritical point M the character of the line should change to first
order. Ground state degeneracies are given in square brackets.
The dashed line indicates the region where strong size effects prevent
us from getting accurate data for the exponents.

\item[Figure 2]
Ground state phase diagram for the SOS model as obtained from
Fig.\ 1 using the coupling relationships between the two models. We have
taken here $K_4=0.1 K_{2y}$.

\item[Figure 3]
Schematic top view of the lattice. The site notation $(i,j)$ is
explicitly indicated. The white and black sublattices are denoted by open
and solid circles, respectively. The dashed line represents the time-slice
set up for performing the spin-chain mapping.
The dimensions $(a_x,a_y)$ of the $(1\times 1)$ unit cell are shown.

\item[Figure 4]
Schematic height profiles, along a time-slice of Fig.\ 3, of the two
ground states of the unreconstructed surface, and of the four missing-row
ground states with their clock label.
The corresponding spin configurations are explicitly indicated.

\item[Figure 5]
(a) Size scaling of the gap to the first $S=1$ excited state.
(b) Size scaling of the gap to the first excited state of smallest momentum.

\item[Figure 6]
Summary of the results for the exponent $\eta$ and $\eta_z$ along
the $J_z=J$ line. A transition, at which $\eta=1/\eta_z\rightarrow 1$, is
clearly seen for $J_2/J\approx 1.2\div 1.3$.

\item[Figure 7]
Size scaling of the $q\rightarrow 0$ limit of the structure factor
for several values of $J_2$ along the $J_z=J$ line.

\item[Figure 8]
Size scaling of the ``specific heat''
$C_v=\partial^2\epsilon_{GS}/\partial J_2^2$ along the $J_z=J$ line.
The peak position saturates at $J_2^{(c)}\approx 1.66J$.
The logarithmic nature ($\alpha=0$) of the divergence of $C_v$ is demonstrated
by the inset.

\item[Figure 9]
Size scaling of the $\pi/2$ component of the spin structure factor
showing the $N^{3/4}$ divergence at the Ising critical point
$J_2^{(c)}\approx 1.66J$.

\item[Figure 10]
Size scaling of the gap to the first excited $S=1$ state in the
dimer phase ($J_2/J=1.6$,$J_z/J=1$). The inset shows that the $S=0$ state
of momentum $\pi$ is degenerate with the ground state for
$N\rightarrow \infty$.

\end{description}

\begin{references}
\bibitem{PBW:NOTA}
$P_{(1\times 1)}$ is denoted by $P_{BW}$ in Ref.\ \cite{Mazzeo}. We prefer
to use the symbol $P_{BW}$ for the order parameter which characterizes the
disordered flat phase (see sections 4 and 5).
\bibitem{NOTA1}
This kind of mapping is slightly different in spirit from the ``fermionic
mappings'' performed in Refs.\ \cite{MdN_92,Bal_Kar}. With the exception of
a ``time continuum limit'', it is essentially exact and one does not need to
include in the quantum hamiltonian extra terms describing the merging of
four steps (the ``fermions'' of a fermionic mapping) at the boundary of
finite terraces on the surface. See also Ref.\ \cite{NOTA2}.
\bibitem{NOTA2}
This Wigner-Jordan spinless fermion should not be confused with the fermions
introduced in Refs.\ \cite{MdN_92,Bal_Kar} (which have a spin and describe
steps on the surface), and is just an equivalent way of writing Pauli spin
operators. In particular, there is no need in Eq.\ \ref{fermion_mod:eqn} for
fermion non-conserving terms describing ``dislocations'', as is the case for
the ``fermionic mappings'' of Refs.\ \cite{MdN_92,Bal_Kar}.
\end{references}
\end{document}